# Knowledge bases over algebraic models. Some notes about informational equivalence

**Abstract**

The recent advances in knowledge base research and the growing importance of effective knowledge management raised an important question of knowledge base equivalence verification. This problem has not been stated earlier, at least in a way that allows speaking about algorithms for verification of informational equivalence, because the informal definition of knowledge bases makes formal solution of this problem impossible.

The goal of this paper is to provide an implementable formal algorithm for knowledge base equivalence verification based on the formal definition of knowledge base given in [24, 26, 28, 29] and to study some important properties of automorphic equivalence of models. We will describe the concept of equivalence and formulate the criterion for the equivalence of knowledge bases defined over finite models. Further we will define multi-models and automorphic equivalence of models and multi-models that are generalization of automorphic equivalence of algebras.

## 1 Introduction and Motivation

The paper is inspired by a natural question:

*When two knowledge bases are equivalent?*

This question contains some uncertainty, namely it operates with the terms "knowledge base" and "equivalence of knowledge bases". Let us dwell briefly on these notions.

### 1.1 Knowledge bases. Descriptive definitions.

As a rule knowledge bases are defined in a various descriptive ways. The definitions reflect a common sense intuition how a knowledge base should look like. They are informal and well known for the specialists in computer science. For the sake of completeness and for the needs of mathematicians looking for applications we provide the reader with some of them.

A *knowledge base* is defined as a special kind of database for knowledge management. It provides the means for the computerized collection, organization, and retrieval of knowledge.

In its turn, *knowledge management* comprises a range of practices used to identify, create, represent, and distribute *knowledge*.

The definition of "knowledge" is equally a philosophical and a practical task. There is no single agreed definition of knowledge presently, and there remain numerous competing theories. In any case knowledge is some essence which requires *representation of knowledge*. Various artificial languages and notations have been proposed for representing knowledge. They are typically based on logic and mathematics, and have easily parsed grammars to ease machine processing [11, 20, 21, etc.].

Knowledge bases store knowledge in a computer-readable form, usually for the purpose of having automated deductive reasoning applied to them. They contain a set of data, often in the form of rules that describe the knowledge in a logically consistent manner. Logical operators, such as conjunction and disjunction, may be used to build knowledge up from the atomic data. Consequently, classical deduction can be used to reason about the knowledge in the knowledge base.

In general, a knowledge base is not a static collection of information (like a database), but a dynamic resource that may itself have the capacity to learn, as part of an artificial intelligence component. These kinds of knowledge bases can suggest solutions to problems sometimes based on feedback provided by the user, and are capable of learning from experience (like an expert system). Knowledge representation, automated reasoning, argumentation and other areas of artificial intelligence are tightly connected with knowledge bases.

## 1.2 Equivalence problem

One can ask, for example, whether *google* and *yahoo* are equivalent? Obviously, we need to restrict concept of equivalence to some special meaning. For example, they are equivalent if they answer in the same time, or they are accessible in the same way, or using fees of these systems are the same, etc., etc., depending on equivalence criterion.

We study an equivalence of knowledge bases in respect to their informational abilities. In other words, we would like to discuss *informational equivalence of knowledge bases*. If we ask *google* and *yahoo* the same question we expect to get the equivalent answers. It means, we expect to get the same information but may be in different formats. Thus, we can specify the main question stated in the beginning of the paper in a more precise form:

When two knowledge bases are informationally equivalent?

The principal task here is to find out whether the problem of informational equivalence verification is algorithmically solvable. If we concentrate on finite objects then the reasonable answer is yes, we can build the step-by-step procedure used to solve the problem. But when we consider infinite objects it may be problematic. Evidently, knowledge bases are the example of this case (for more details see subsections 2.2 and 4.2). On other side, if we could find some finite invariant (or system of invariants) of a knowledge base, such that equivalence of those invariants would involve equivalence of corresponding knowledge bases, then the problem would turn to algorithmically solvable.

**Example:** Let us consider two vector spaces: plane $R^2$ and tree-dimensional space $R^3$. These two objects are infinite. Whether they are equivalent? Intuitively we can answer no. But the precise answer follows from the fact, that dimension of the space is its invariant. In our example dimensions of two spaces are not the same: two and tree. So the answer is no, these objects are not equivalent.

### 1.3 Problem in question and the main results

The problem of equivalence in database research aroused already in 80th of the previous century. Beniaminov [5], Beeri-Mendelzon-Sagiv-Ullman [4] and others proposed algorithms for verification of databases equivalence using database schemes.

In [4] the authors introduced the notion of a fixed point of a database scheme. In this setting two relational database schemes are equivalent if their sets of fixed points coincide. Correspondingly, two relational databases are equivalent if their sets of all fixed points intersected with the sets of feasible instances coincide (see [4] for details).

This approach was based on comparing database structure and did not consider its content, thus, is was more suitable for evaluation of structural rather than informational equivalence. B. I. Plotkin proposed a mathematical model of a database [26] and gave a formal definition of the database informational equivalence concept based on this model [25]. Further, on the ground of this definition the problem of databases equivalence verification was considered [27].

Knowledge base systems go beyond the relational model toward handling complex data that may include rules. They combine features of database management systems with artificial intelligence techniques. The existing knowledge bases are defined using informal description of internal relationships and as a consequence they do not allow to identify equivalent knowledge represented in different ways by different knowledge base implementations. This problem can be solved only by providing a formal mathematical model of a knowledge base. Such model was presented in [25]. In the series of papers [23, 24, 26, 27, 28, 29], the authors also proposed a solution for the knowledge base equivalence problem using an algebraic geometry approach, category theory, graph theory and group-theoretic methods [2].

In particular they proved that :

**Theorem**. *Two knowledge bases are informationally equivalent if and only if the corresponding subjects of knowledge are automorphically equivalent.*

This result introduces the notion of automorphic equivalence as a key tool of the theory. Study of this notion is one of the main objectives of this paper. We show that this notion is much wider than the notion of isomorphism. This means that two knowledge bases which are far from being isomorphic can be informationally equivalent. We prove also that the problem of informational equivalence of knowledge bases is algorithmically solvable in case of finite subjects of knowledge (see Theorem).

We provide an implementable formal algorithm for knowledge bases equivalence verification based on the formal definition of a knowledge base given in [26]. We will describe the concept of equivalence and formulate the criterion for the equivalence of knowledge bases defined over finite models. Further we will define multi-models and automorphic equivalence of models and multi-models that are generalizations of automorphic equivalence of algebras.

We hope that the ability to verify informational equivalence of two different knowledge bases can be used to increase efficiency of knowledge retrieval and detection of hidden knowledge. If retrieving information from one knowledge base may be problematic, the same information can be possibly easily accessible in another informationally equivalent knowledge base. Another application of knowledge base equivalence verification is the disambiguation of information that arrived from different sources or was encoded in different formats. In this case information that is considered equivalent can be skipped.

The paper is organized as follows.

Sections 2-3-4 are devoted to the notion of automorphic equivalence. Section 2 provides algebraic background. Here we recall notations of algebraic model and multi-model and automorphic equivalence of algebras, models and multi-models. Section 3 defines two ways to build automorphically equivalent multi-model to a given model. Section 4 discusses some properties of automorphic equivalence of multi-models, like graph tree structure preservation and preservation of graph connectedness.

Sections 5 and 6 introduces algebraic model of knowledge bases and their informational equivalence according to [25].

Section 7 is the algorithm outline for verification of knowledge base informational equivalence.

Finally, section 8 provides our conclusions.

## 2. Algebraic Background

In this section we will discuss notions of model, multi-model and automorphic equivalence of models, multi-models and algebras.

### 2.1. Automorphic Equivalence of Algebras

We will use the following *notation for algebra*: $\Psi = (A, \Omega, f)$, where $A$ is a set of elements, $\Omega$ is a set of operations on $A$ and $f$ is an interpretation of these operations.

A one-to-one correspondence between two algebras preserving all operations is called isomorphism of algebras. An isomorphism from $\Psi$ onto $\Psi$ is called an *automorphism* of $\Psi$. The group of all automorphisms of $\Psi$ will be denoted by $Aut(\Psi)$. When we concentrate on some interpretation $f$ of algebra operations $A$, we mean that automorphisms of algebra will transform this interpretation to itself. We will denote this group of automorphisms by $Aut(f)$.

**Definition [26]**. *Let us consider two algebras $(A, \Omega_1, f_1)$ and $(B, \Omega_2, f_2)$ that have groups of automorphisms $Aut(f_1)$ and $Aut(f_2)$ accordingly. We call these algebras automorphically equivalent if there exists a bijection $\delta: A \to B$ such that the groups of automorphisms are conjugated by $\delta$. In other words the following equality*

$$Aut(f_2) = \delta \, Aut(f_1) \, \delta^{-1}$$

*holds.*

It is clear that if algebras are isomorphic then they are also automorphically equivalent. However, the opposite assertion is not always correct.

## 2.2. Models

**Definition**. *We define a model as a triple $(D, \Phi, f)$ where $D$ is a data domain, that is, an algebra in a variety of algebras $\Theta$ (for example, vector space over a field), $\Phi$ is a set of symbols of relations, $f$ is one of possible interpretations of these symbols as real relations in D, i. e., if $\phi \in \Phi$ is an n-ary relation in $\Phi$, then $f(\phi)$ is a subset of the Cartesian product $D^n$. Moreover, $D$ may be a multi-sorted set, i. e., $D = \{D_i, i \in \Gamma\}$, where $\Gamma$ is a set of sorts.*

For example, consider a group of students. The multi-sorted domain $D$ is defined as $D = \{D_1, D_2, D_3\}$, where $D_1$ is a set of students, $D_2$ is a set of subjects and $D_3$ is a set of all possible grades, $\Gamma = \{1, 2, 3\}$. On this three-sorted set $D$ one may consider ternary relation $\phi(x, y, z)$ : x is a student, y is a subject and z is a grade. We say that this relation holds on the given set $D$ if student x got grade z on subject y. This is interpretation $f$ of the symbol $\phi$. In multi-sorted algebra we consider the type of relation and the type of operations instead of arity of relation and arity of operations like in one-sorted case. The set $D$ is considered together with algebraic operations on it. Usually operations satisfy some algebraic laws such as the laws of a Boolean algebra or semi-group laws.

## 2.3. Multi-Models

**Definition**. *A multi- model is a triple* $(D, \Phi, F)$, *where* $D$ *is a data domain (an algebra)*, $\Phi$ *is a set of symbols of relations*, $F$ *is a set of different interpretations of* $\Phi$ *on* $D$.

A model $(D, \Phi, f)$ is a particular case of a multi-model $(D, \Phi, F)$. The definition of multi-model takes into account that the instance (interpretation) $f$ can change, for example over time or under some other circumstances. All these $f$ constitute the set $F$. In general multi-models may be infinite but we consider only the finite ones.

## 2.4. Automorphic Equivalence of Models and Multi-Models

For the given model $(D, \Phi, f)$ we have a group $Aut(f)$ consisting of all bijections $s: D \to D$ compatible with the interpretation of symbols of relations. This means that for every n-ary relation $\phi \in \Phi$ and every element $(a_1, a_2, ..., a_n) \in f(\phi)$ the element $(sa_1, sa_2, ..., sa_n)$ belongs to $f(\phi)$ as well. In the case of a multi-sorted model $s = (s_i, i \in \Gamma)$. The set of all such $s$ form the group of automorphisms of the model denoted by $Aut(f)$ (it should be noted that there are some cases when this group is trivial).

Recall that two models $(A, \Phi_1, f_1)$, $(B, \Phi_2, f_2)$ are called isomorphic if the sets $\Phi_1$ and $\Phi_2$ coincide and there is a bijection $\sigma: A \to B$ which is an isomorphism of algebras and for any n-ary relation $\phi \in \Phi$ we have $(a_1, a_2, ..., a_n) \in f_1(\phi)$ if and only if $(\sigma(a_1), \sigma(a_2), ..., \sigma(a_n)) \in f_2(\phi)$.

**Definition.** *Let us consider two models* $(A, \Phi_1, f_1)$ *and* $(B, \Phi_2, f_2)$. *Assume that* $A$ *and* $B$ *are algebras with the same operations, defined by the variety of algebras* $\Theta$. *They are called automorphically equivalent, if there is an isomorphism* $\mu: A \to B$ *such that groups of automorphisms are conjugated by this isomorphism, i.e.,*
$Aut(f_2) = \mu \, Aut(f_1) \mu^{-1}$.

The difference between automorphic equivalence of models and automorphic equivalence of algebras is the transformation $\mu: A \to B$. For models we use the notion of isomorphism of algebras $\mu$ while for algebras we referred to it as a bijection of sets $\delta$.

**Definition .** *Two multi-models $(A, \Phi_1, F_1)$ and $(B, \Phi_2, F_2)$ are called automorphically equivalent, if there is a bijection $\alpha: F_1 \to F_2$, such that the models $(A, \Phi_1, f)$ and $(B, \Phi_2, f^\alpha)$ are automorphically equivalent for every $f \in F_1$.*

This means that it is possible to correlate the instances of these multi-models in such a way that the corresponding models turn to be automorphically equivalent.

## 3. Multi-Models Automorphically Equivalent to a given one

Let us consider two ways to construct a multi-model $(A, \Phi, F_1)$, which is automorphically equivalent to the given multi-model $(A, \Phi, F)$. We will use the multi-model to investigate automorphic equivalence properties (section 4).

We will investigate two possible approaches:
1. In order to get $f' \in F_1$ some transformation $\sigma$ will be applied to $f \in F$.
2. We will define $f' \in F_1$ as a complement to $f \in F$.

Let us look at these two cases in more detail.

### 3.1. Construction of Automorphically Equivalent Multi-Models Using Transformation $\sigma$

Let $\sigma$ be an automorphism of algebra A. For every interpretation $f \in F$ we will construct another interpretation $f^\sigma$ by the following rule: for n-ary relation $\varphi \in \Phi$ and row $(a_1, a_2, ..., a_n) \in A$, we set: $(a_1, a_2, ..., a_n) \in f^\sigma(\varphi)$ if and only if $(a_1^{\sigma^{-1}}, a_2^{\sigma^{-1}}, ..., a_n^{\sigma^{-1}}) \in f(\varphi)$.

Now we build a mapping $\mu: A \to A$, such that $\mu(a) = \sigma(a)$ for every $a \in A$, and $\alpha: F \to F_1$, such that $f^\alpha = f^\sigma$ for every $f \in F$.

Here we define a multi-model $(A, \Phi, F_1)$ which is isomorphic to the given multi-model $(A, \Phi, F)$. Moreover, $Aut(f^{\alpha}) = \sigma Aut(f) \sigma^{-1}$ which means that these two multi-models are automorphically equivalent.

Thus, any automorphism of the algebra $A$ induces a multi-model which is isomorphic and, consequently, automorphically equivalent to the given one.

## 3.2. Construction of Automorphically Equivalent Multi-Models Using $\overline{f}$

Let $(A, \Phi, F)$ be a multi-model. For every $f \in F$ we build $\overline{f}$ using the following rule: $\overline{f}(\varphi) = \overline{f(\varphi)}$, where "bar" denotes the complement in the corresponding Cartesian product. Denote by $\overline{F}$ the set of all $\overline{f}$. Let us consider a mapping $\alpha: F \to \overline{F}$ defined by $f^{\alpha} = \overline{f}$. It is clear that this map is a bijection. Now we use an identity transformation mapping $\mu: A \to A$. Obviously, here $Aut(f) = Aut(\overline{f})$ and the equality $Aut(\overline{f}) = \mu\, Aut(f) \mu^{-1}$, where $f \in F$, takes place.

Thus, the multi-models $(A, \Phi, F)$ and $(A, \Phi, \overline{F})$ are automorphically equivalent but they are not isomorphic.

## 4. Some Properties of Automorphic Equivalence

Our next aim is to investigate some properties of automorphic equivalence. With this end we consider graphs as a particular example of models.

## 4.1. Graphs

**Definition.** A g*raph is a pair of sets* $G = (V, E)$ *where* $V$ *is a set of vertices (points) and* $E$ *is a set of edges (pairs of points, connected by the edges).*

To each graph $G = (V, E)$ corresponds a model $(V, \varphi, E)$ where $V$ is a domain of the model, $\varphi$ is the only relation that exists on the graph and defines edges between vertices, $E$ is an interpretation of the relation $\varphi$ on the domain $V$, i.e., $E \subseteq V \times V$.

As usual, an automorphism of a graph is a permutation on the set of vertices preserving edges.

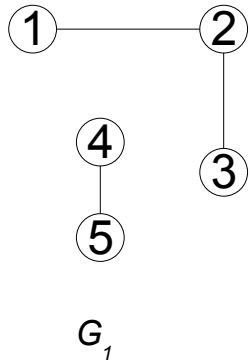 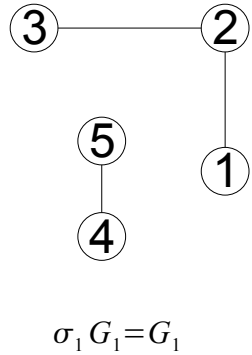 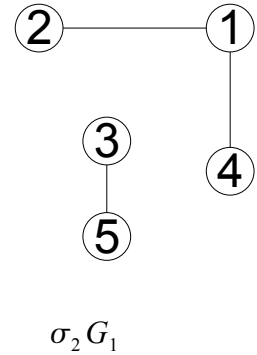

$G_1$          $\sigma_1 G_1 = G_1$          $\sigma_2 G_1$

**Figure 1.**          **Figure 2.**          **Figure 3.**

For example, permutation $\sigma_1 = \begin{pmatrix} 1 & 2 & 3 & 4 & 5 \\ 3 & 2 & 1 & 5 & 4 \end{pmatrix}$ is an automorphism of the graph $G_1$ since $\sigma_1 G_1 = G_1$. If we apply another permutation, for example, $\sigma_2 = \begin{pmatrix} 1 & 2 & 3 & 4 & 5 \\ 2 & 1 & 4 & 3 & 5 \end{pmatrix}$, we get a graph that is isomorphic but not identical to $G_1$.

All automorphisms of a graph constitute a group, which is a subgroup of symmetric group acting on vertices. The automorphism group of a graph characterizes its symmetries, and, therefore, is very useful in determining some of its properties.

### 4.2. Investigation of a Tree Structure Preservation

In graph theory a tree is a graph in which any two vertices are connected by exactly one path. Alternatively, any connected graph with no cycles is a tree. We show that automorphic equivalence of graphs does not preserve tree structure of a graph.

### 4.2.1. Building automorphically equivalent Multi-Models Using Algebra Automorphism

Assume that $G_1 = (V_1, E_1)$ is a tree and $G_2 = (V_2, E_2)$ is an arbitrary graph. Let $G_1$ and $G_2$ be automorphically equivalent graphs. It means that there exists a bijection $\alpha$ that transforms $E_1$ to $E_2$, a bijection $\mu$ that transforms $V_1$ to $V_2$ and $Aut(G_1)$ and $Aut(G_2)$ are conjugated: $Aut(G_2)=\mu\, Aut(G_1)\mu^{-1}$.

Now let us consider the following example. We have two graphs on Figure 7:

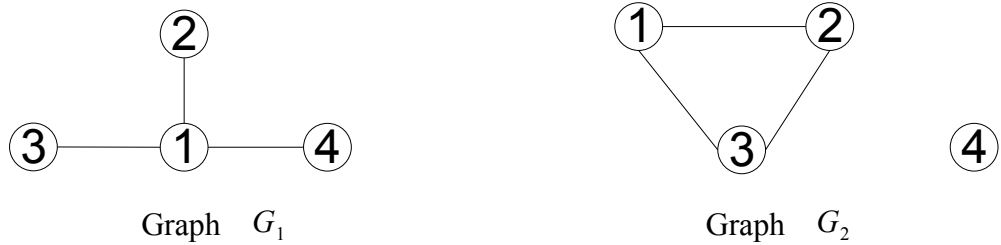

Graph $G_1$    Graph $G_2$

**Figure 7**

The set of vertices for graph $G_1$ is $V_1=(1,2,3,4)$ and set of edges is
$E_1=(e_1^1=(1,2), e_1^2=(2,1), e_1^3=(1,3), e_1^4=(3,1), e_1^5=(1,4), e_1^6=(4,1))$. The automorphisms group consists of all permutations of 2, 3 and 4.

For graph $G_2$ we have $V_2=(1,2,3,4)$ and the set of edges is
$E_2=(e_2^1=(1,2), e_2^2=(2,1), e_2^3=(1,3), e_2^4=(3,1), e_2^5=(2,3), e_2^6=(3,2))$. The automorphisms group consists of all permutations of 1, 2 and 3.

Let us demonstrate that these two graphs are automorphically equivalent.

1. There exists a bijection $\alpha: E_1 \to E_2$

$$\alpha = \begin{pmatrix} e_1^1 & e_1^2 & e_1^3 & e_1^4 & e_1^5 & e_1^6 \\ e_2^1 & e_2^2 & e_2^3 & e_2^4 & e_2^5 & e_2^6 \end{pmatrix}$$

2. There exists a bijection $\mu$ as defined below (here we use cyclic representation of permutations):

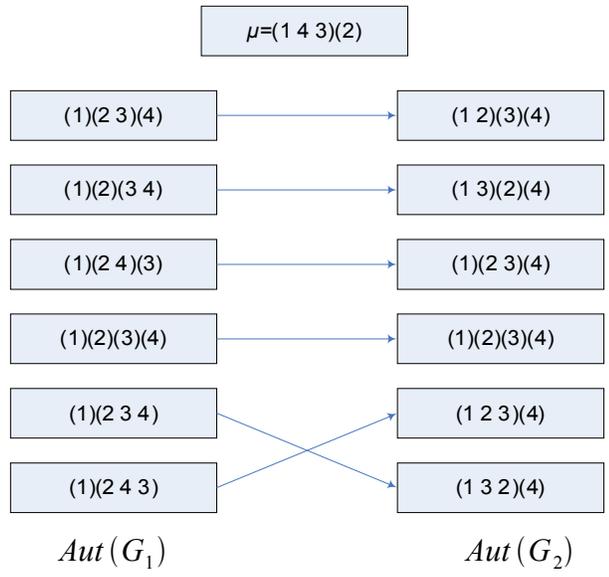

Groups of automorphisms are conjugated by bijection $\mu$. Therefore, graphs are automorphically equivalent.

This example illustrates that automorphic equivalence of two graphs does not preserve the basic characteristics of those graphs, like "being a tree".

Let us consider an additional example with directed graphs (Figure 8).

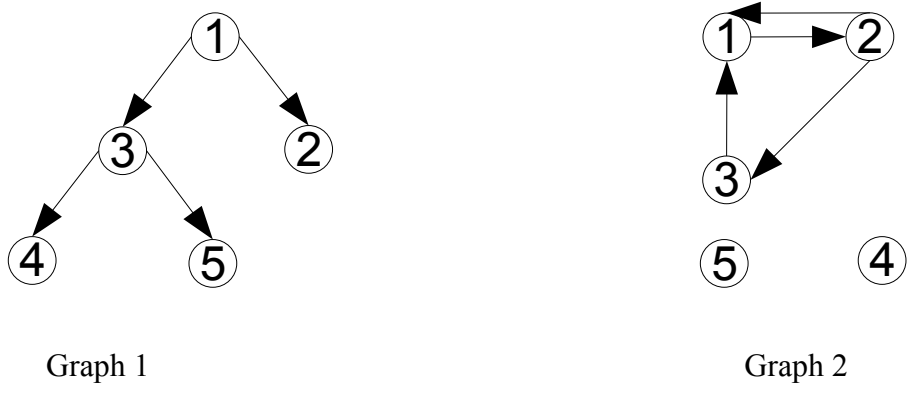

           Graph 1                                        Graph 2

**Figure 8.**

The set of vertices for graph $G_1$ is $V_1=(1,2,3,4,5)$ and the set of edges is $E_1=(e_1^1=(1,2), e_1^2=(1,3), e_1^3=(2,4), e_1^4=(2,5))$. The automorphisms group consists of the following permutations $\begin{pmatrix} 1 2 3 4 5 \\ 1 2 3 5 4 \end{pmatrix}, \begin{pmatrix} 1 2 3 4 5 \\ 1 2 3 4 5 \end{pmatrix}.$

The vertices of graph $G_2$ are $V_2=(1,2,3,4,5)$ and edges of this graph are $E_2=(e_2^1=(1,2), e_2^2=(2,3), e_2^3=(3,1), e_2^4=(2,1))$. The automorphisms of the graph are

$\begin{pmatrix} 1 2 3 4 5 \\ 1 2 3 5 4 \end{pmatrix}, \begin{pmatrix} 1 2 3 4 5 \\ 1 2 3 4 5 \end{pmatrix}.$

We can see that these two graphs are automorphically equivalent.

1. There exists bijection $\alpha : E_1 \to E_2$

$$\alpha = \begin{pmatrix} e_1^1 & e_1^2 & e_1^3 & e_1^4 & e_1^5 & e_1^6 \\ e_2^1 & e_2^2 & e_2^3 & e_2^4 & e_2^5 & e_2^6 \end{pmatrix}$$

2. There exists bijection $\mu$ as defined below (here we use cyclic representation of permutations):

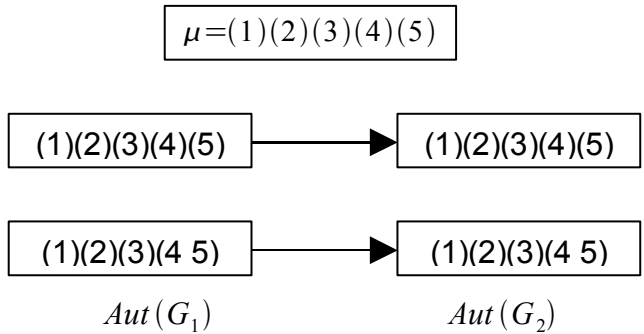

Groups of automorphisms are conjugated by bijection $\mu$. Therefore, graphs are automorphically equivalent. Thus, the graphs on Figure 8 are automorphically equivalent but the property of being a tree is not preserved since $G_2$ is not a tree.

### 4.2.2. Constructing Automorphically Equivalent Multi-Models Using Complement interpretation

For multi-model of the given graph $(A, \Phi, F) = (V, \Phi, E)$ we will build an automorphically equivalent multi-model $(A, \Phi, \overline{F}) = (V, \Phi, \overline{E})$ using the approach from Section 3.2..

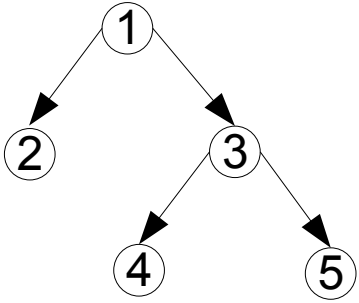 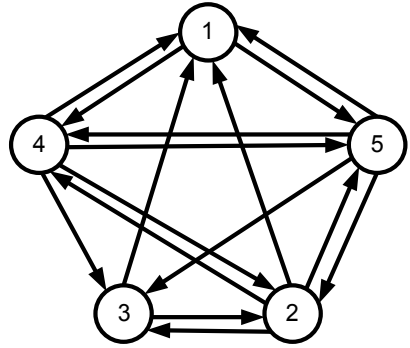

**Figure 9**. Given graph $G=(V,E)$     **Figure 10**. Graph $\overline{G}=(V,\overline{E})$

In order to build the second graph we connect two vertices that were not connected in the original graph and vice versa – if vertices were connected in the original graph then we do not connect them in the new one.

Here we have just one interpretation in each multi-model, so there is no problem to build a bijection $\alpha: F \rightarrow \overline{F}$. This bijection is $\alpha(E)=\overline{E}$.

The second bijection $\mu: A \rightarrow A$ is the identity transformation and in this example it is the permutation $\mu=\begin{pmatrix} 1,2,3,4,5 \\ 1,2,3,4,5 \end{pmatrix}$. According to the criterion of automorphic equivalence two our multi-models are automorphically equivalent.

Since the second graph is not a tree, automorphic equivalence does not preserve this property.

## 4.3. Graph Connectedness and Automorphic Equivalence.

### 4.3.1. Building automorphically equivalent Multi-Models Using Algebra Automorphism

From the example in subsection 4.1.1, we can see that the first graph is connected (tree) but the second one is not. As we proved already two corresponding multi-models are automorphically equivalent. So, automorphic equivalence of multi-models does not keep connectedness of the graphs.

## 4.3.2. Building automorphically equivalent Multi-Models Using Inverse interpretation

Again we start from multi-model of connected graph $(A, \Phi, F) = (V, \Phi, E)$ and we build automorphically equivalent multi-model by the second approach, $(A, \Phi, \overline{F}) = (V, \Phi, \overline{E})$.

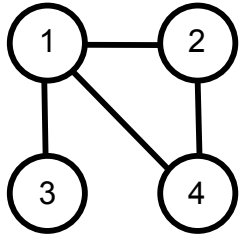 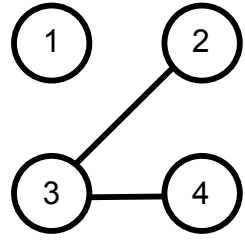

**Figure 11**. Given graph $G=(V,E)$      **Figure 12**. Graph $\overline{G}=(V,\overline{E})$

Here we use bijection $\alpha: F \rightarrow \overline{F}$, where $\alpha(E) = \overline{E}$, and identical bijection $\mu: A \rightarrow A$, i. e., $\mu = \begin{pmatrix} 1,2,3,4 \\ 1,2,3,4 \end{pmatrix}$. Then the following equality takes place $Aut(\overline{f}) = \mu \, Aut(f) \mu^{-1}$. This shows that two multi-models are automorphically equivalent.

In this second example we also see that automorphic equivalence does not save connectedness of the graph.

## 5. Algebraic Model of a Knowledge Base

According to the formal mathematical definition of a knowledge base, suggested in [25], knowledge base is described using three components: t*he syntax of knowledge representation, knowledge subject and knowledge content,* all of which are considered further.

    *1.    Knowledge Metadata (Syntax of Knowledge Representation)*

Knowledge metadata is described by set of formulas $T$ using only FOL (First Order Logic). Two FOL formulas are considered to be equivalent if they return the same content for all models. This transition from FOL formulas to classes of equivalent formulas leads us to algebraic logic (AL). The corresponding AL in this case is a Boolean algebra with quantifiers and equalities. Such approach enhances the ability to describe the knowledge. The same knowledge may be defined by different descriptions (where one description may contain the other). In this sense some maximal description of knowledge exists (i.e., the one that is not contained in any other description). This maximal description has some algebraic features.

The category of knowledge description is denoted by $LD_{\Phi\Theta}$ (Logical Description). This logical description depends on $\Phi$, the set of symbols of relations in model $(D, \Phi, f)$, and $\Theta$, the variety of algebras to which $D$ belongs.

## 2. *Knowledge Domain (Subject of Knowledge)*

Knowledge domain or subject of knowledge is described by a model $(D, \Phi, f)$ as explained in subsection 2.2. A knowledge base is called *finite* if the corresponding data algebra $D$ is finite

## 3. *Knowledge Meaning (Content of Knowledge)*

To every description of knowledge, i.e., to the set of formulas $T$, and a subject of knowledge $(D, \Phi, f)$ corresponds a content of knowledge denoted by $T^f$ ($T$ from the description and $f$ from the subject). We consider formulas of $T$ to be equations. Then $T^f$ is a locus of the points satisfying the system of equations $T$. This locus lies in some affine space $D^n$ (like in geometry a system of equations can be solved on a plane, in a three-dimensional space, etc.). In algebraic geometry such locuses are called algebraic sets or algebraic varieties. There are relations (links) between different locuses, and, thus, we can define a category of algebraic sets. Content of knowledge is an object of this category denoted by $CK_{\Phi\Theta}(f)$. Like the previous category this one depends on $\Phi$ and $\Theta$ too, but in addition it takes into account $f$, so that we have for

$CK_{\Phi\Theta}(f)$ every $f \in F$.

Additionally there exist functors $Ct_f$ from $LD_{\Phi\Theta}$ to $CK_{\Phi\Theta}(f)$, that match syntax to the corresponding content.

Knowledge bases are formally defined by category $LD_{\Phi\Theta}$ and by set of functors $Ct_f$, $f \in F$, to categories $CK_{\Phi\Theta}(f)$ (here there is only one logical description category and many functors to content categories depending on state $f$ of the knowledge base).

Knowledge base depends on the chosen multi-model. We denote it by $KB(D,\Phi,F)$.

## 6. Informational Equivalence of Knowledge Bases

Informally two knowledge bases are called equivalent if all information that can be retrieved from the first knowledge base can be also retrieved from the second one and vice versa.

The formal definition of knowledge base equivalence follows [25].

Let us consider two multi-models $(A,\Phi_1,F_1)$ and $(A,\Phi_2,F_2)$ with the corresponding knowledge bases $KB_1$ and $KB_2$. Each of these knowledge bases has one logic description category and several knowledge content categories. The relation between these categories is represented on the following diagrams (Figure 12 and Figure 13):

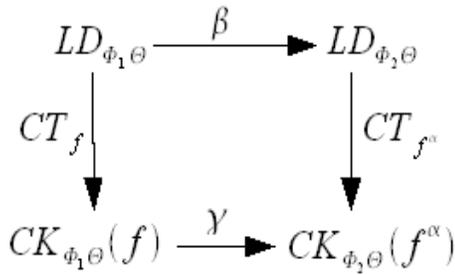 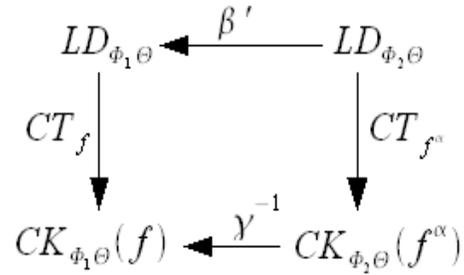

Figure 12.          Figure 13.

Here

- $\alpha: F_1 \rightarrow F_2$ is a bijection between sets of states $f$,
- $\beta, \beta'$ are homomorphisms of knowledge description categories in opposite directions,
- $\gamma$ is an isomorphism of knowledge content categories.

**Definition ([25])**. *Knowledge bases $KB_1$ and $KB_2$ are called informationally equivalent if and only if it is possible to choose $\alpha, \beta, \beta', \gamma$ that match the two commutative diagrams presented on Figure 12 and Figure 13.*

This definition of informational equivalence is not quite suitable for the purpose of practical implementation. However, the key result of [25] says that there exists a further reduction from knowledge bases to multi-models and their automorphic equivalence. Namely,

**Theorem ([25]):** *Knowledge bases $KB$ $(A, \Phi_1, F_1)$ and $KB$ $(B, \Phi_2, F_2)$ are informationally equivalent if and only if their finite multi-models are automorphically equivalent.*

*In other words two knowledge bases are informationally equivalent if and only if the corresponding subjects of knowledge are automorphically equivalent.*

This theorem provides a possibility to build an exact algorithm for knowledge bases informational equivalence verification.

## 7. Knowledge Bases Informational Equivalence Verification Algorithm Outline

In section 6 we investigated theorem for reduction of the problem of infinite knowledge bases informational equivalence verification to the problem of finite multi-models automorphic equivalence checking [25]. We also saw that there ensues very important *corollary* from this theorem:

**Theorem:** For two given multi-models $(A, \Phi_1, F_1)$ and $(B, \Phi_2, F_2)$ there exists algorithm for their automorphic equivalence verification.

We *prove* this corollary by constructing formal algorithm:

if algebras $A$ and $B$ are not isomorphic

        exit with "**not automorphically equivalent**"

    else

        find group of automorphisms $Aut(A)$

        find group of automorphisms $Aut(B)$

        find set $Aut_{FA}$ of all subgroups $Aut(f)$ from $Aut(A)$, where $f \in F_1$

        find set $Aut_{FB}$ of all subgroups $Aut(f)$ from $Aut(B)$, where $f \in F_2$

        build bipartite graph $(V, E)$ of conjugated $(f_i, f_j)$ from $F_1 x F_2$

        if exists $f_i$ without corresponding $f_j$

            exit with "**not automorphically equivalent**"

        else

            find bijection $\alpha: F_1 \rightarrow F_2 \,|\, (f, f^\alpha) \in E$ for all $f \in F_1$

            if $\alpha$ exists

                exit with "**automorphically equivalent**"

            else

                exit with "**not automorphically equivalent**"

## 8. Conclusion

We have presented two approaches to building multi-model automorphically equivalent to a given one. The first approach uses algebra automorphism and the second one uses inverse interpretation of relation names.

We studied some properties of automorphic equivalence of multi-models using graphs as an example. We have shown that it is possible to provide a semantic mapping between the terms of graph theory and terms that we used to describe our model. We also proved that automorphic equivalence of graphs does not preserve their connectedness or tree structure.

Finally we provided an outline for a practically implementable algorithm for automorphic equivalence verification of multi-models.